\documentclass{article}
\usepackage{graphicx} 
\usepackage{url}
\begin{document}
\title{An optimal MOO strategy\footnote{
  This document is an English translation of a paper ``Tetsuro Tanaka, An Optimal MOO Strategy, Game Programming Workshop - Japan (1996)'' written in Japanese. Thanks to grammarly.com and deepl.com. }}
\author{
Tetsuro Tanaka, Faculty of Engineering, University of Tokyo\\
{\tt tanaka@ipl.t.u-tokyo.ac.jp}\footnote{The affiliations and e-mail addresses of the authors listed here are as of 1996. The affiliation and e-mail address of the author
as of 2022 is "Information Technology Center, The University of Tokyo," ktanaka@g.ecc.u-tokyo.ac.jp.}
}
\date{}
\maketitle
\begin{abstract}
We calculated a fixed strategy that minimizes the average number of guesses (minimum strategy) for the number-guessing game MOO by exhaustive search. 
Although the minimum strategy for a similar game, mastermind, has been reported by \cite{kenji}, this study seems to be the first to find the minimum strategy for MOO with a larger search space.

When two players play against each other in MOO, the minimum strategy is not always the strongest fixed strategy. First, we compute a fixed strategy that has the maximum winning rate when played against the minimum strategy. Then we confirm that there is no fixed strategy with a winning rate exceeding 0.5 against this strategy. This result shows that MOO is a game with the strongest fixed strategy.

\end{abstract}
\section{Rules of MOO}

MOO is a popular number-guessing game that has been called by various names such as Hit \& Blow, Cow \& Bull.\footnote{There is also a MUD (Multi-User Dungeon) game of the same name.} The rules of the game are as follows. 

\begin{itemize}
\item Two players write on a piece of paper a four-digit number (hereafter called a secret MOO number) consisting of numbers different from each other and keep it out of sight of both players. The first digit may be a zero, as in 0586. 

\item Each player shows a MOO number to guess the opponent's secret MOO number. The opponent responds with the number of bulls, whose locations and numbers are the same, and cows, whose numbers are the same, but their locations differ from the secret MOO numbers. Calculating the number of bulls and cows from two MOO numbers is called the {\it MOO product} below. You will get four bulls if you guess the opponent's secret MOO number.

\item The player who gets four bulls with fewer guesses wins. If the number of guesses is equal, the game is a tie.
\end{itemize}

Figure \ref{game-example} shows an example of game sequence viewed from only one side, with the guesser and the respondent fixed.
\begin{figure}[htbp]
\begin{center}
\begin{tabular}{rrr}
turn &guess & response \\ \hline
 1  &0123 &2C \\
  2  &1245  &2C \\
  3  &2671  &1B \\
  4  &2850  &1B \\
  5  &9351  &2B2C \\
 6  &3951  &4B \\
\end{tabular}
\end{center}
\caption{An example of game sequence. The secret MOO number is 3951.}
\label{game-example}
\end{figure}

There is a game similar to MOO called Mastermind. Mastermind differs from MOO in that it allows duplicate numbers (colors), and the number of colors is usually only six. Some BSD-like Unixes come with MOO programs as standard equipment.
\section{Related Works}

Programs to generate random MOO numbers and to answer the human guesses with bull and cow numbers have been developed since the early days of MOO. \cite{software} introduces a Cambridge University system developed in 1971.
 The system ranked players according to the average number of guesses they asked and displayed their high scores. The system became very popular among users, and some user even predicted the next guess by analyzing a random number generation algorithm.

Programs that submit MOO problems are also good exercises for programming, and some of them, such as \cite{sympo}, have been the subject program of {\it Programming Symposiums}\footnote{The {\it Programming Symposium} is the name of a Japan Domestic Conference.}, while others, such as \cite{shell}, have been written in shell.

On the other hand, a study of playing MOOs on a computer is also introduced in \cite{software}. There are 5040 (${}_{10}P_{4}$) MOO numbers in total. The key point to making a strong MOO playing program is how to find the best guess among the 5040 MOO numbers for the set of MOO numbers which are obtained by asking some guesses.

For a set of MOO numbers $\pi$, when a MOO number $T$ is chosen as a guess, there are 14 types of responses, 4B, 3B, 2B2C, 2B1C, 2B, 1B3C, 1B2C, 1B1C, 1B, 4C, 3C, 2C, 1C, 0C, and $\pi$ is divided into $t_1,... t_{14}$. We can construct a function $f(T)$ to evaluate this partitioning and minimize it.

J. Larmouth, using the fact that the number of guesses for a set of size $n$ can be estimated as ${\bf log}(n)$, chose the function
$$ f(T)=\sum_{i\in [1,14],t_i\neq \emptyset} |t_i|{\bf log}(|t_i|) (-2log2 ~~{\bf if}~~T\in \pi), $$
and achieved an average number of guesses of 5.24.
B. Landy proposed the function.
$$ f(T)=max(|t_i|) $$
$$ f(T)=\sum_{i=1}^{14} |t_i|{\bf log}(1+|t_i|) $$
$$ f(T)=\sum_{i=1}^{14} |t_i|F(|t_i|) ~~{\bf where }~~F(n)~~ {\bf is~~ the~~ solution~~ of~~ }~~x^x=n.$$

In 1987, this problem was given as an assignment in the Nanopico class of the magazine bit\footnote{The journal bit here refers to a general computer science journal published by the Japanese publisher Kyoritsu Shuppan.}. At that time, the most common solution was to construct $f(T)$ and choose the best guess for each set. The winner's program achieved an average number of guesses of 5.22 (total number of guesses: 26347) using a modified version of B. Landy's formula. 

\section{Computing Minimum Strategies by Exhaustive Search}

When computing $f(T)$, if we consider even the partition of the set when 5040 guesses are asked for each of the partitioned sets $t_1,...,t_{14}$, we can choose better guesses, although the computational complexity increases. This can be thought of as a game tree search problem of depth 2. 

Extending this further, we find that for the set of 5040 MOO numbers before the first guess, the average number of guesses can be minimized by searching the game tree with unlimited depth until the size of each becomes less than or equal to 1. Once this search is performed and the best guesses in each phase are listed in a table, the program to play the game becomes very simple.

Although the possibility of creating a minimum guess table by this exhaustive search was also suggested in \cite{software}, it was considered too difficult at that time, as shown below.
\begin{quote}
 This is far too expensive on computation.
\end{quote}

With the subsequent progress of computers, the minimum strategy for the mastermind was obtained using a perfect search. 
However, the number of guesses in each phase is $6^4 = 1296$ for the mastermind game, whereas it is ${}_{10}P_{4} = 5040$ for MOO. 
Because the number of branches in the search tree is about four times larger and the average number of guesses is larger, it is considered to take more computation time. This is considered to be more computationally time-consuming.

In this study, we paid special attention to the following points in programming. 
\begin{enumerate}
\item Adoption of data structures suitable for high speed
\item Elimination of symmetric guesses
\item Creating a good heuristic function by preprocessing
\end{enumerate}
As a result, using a workstation (Sparc Station 20), we obtained the minimum strategy (total number of guesses: 26274) in about 60 hours for the preprocessing and 30 hours for the calculation of the minimum strategy. The following section describes the details of the method. 

\section{Faster tree search execution}

\subsection{data representation}

We represent MOO numbers as 26-bit integers. The lower 16 bits represent a 4-digit decimal number, with every 4 bits representing one decimal digit. The upper 10 bits represent a mask of digits (from 0 to 9) contained in the MOO number.

To find the bull number of MOO numbers a and b expressed in this way, we can take the lower 16 bits of \verb|a ^ b| (where \verb|^| is exclusive disjunction) and count the number of digits of 0 in decimal. For this purpose, it is sufficient to prepare a table of 64K bytes. To find the number of cows, count the number of bits set to 1 in the upper 10 bits of the table \verb|a & b| and subtract the number of bulls from it. We can calculate them very fast with a table of 1K bytes. The calculation of the MOO product based on this data representation is as follows.
\begin{verbatim}
typedef int Question2;
int moo(Question2 q1,Question2 q2)
{
  int bull,cow;

  bull=bulltable[(q1^q2)&0xffff];
  cow=cowtable[(q1&q2)>>16];
  return(moo_product[bull][cow]);
}
\end{verbatim}

We checked the execution time on Sparc Station 20 and found that it was about three times faster than the naive one. If a table of 5040x5040 size is made, the table can be referred to only once, but since one table requires 25M entries (12.5M bytes if each entry is represented by 4 bits), we did not use this method this time. 

\subsection{Elimination of Equivalent guesses}

Since it is often futile to compute the best guess for a set for all 5040 possible guesses, a check is made to avoid equivalence among the 5040 possible guesses. Guesses that the following iterations of substitution can reach are equivalence guesses.

\begin{itemize}
\item Numbers that do not appear in the set can be substituted. 
(For example, the numbers [0,7,8,9] in the guess for the set [1234,3456,1256] are equivalent)
\item Numbers that do not appear in previous guesses can be substituted.
(For example, if the first guess is 0123 and the second is 1234, then [5,6,7,8,9] can be considered the same in the next guess)
\item A positional and numeric substitution may make from a sequence of guesses to the same sequence of guesses. When there exists such a substitution, the next guess and the guess applied to the substitution are considered the same.
\end{itemize}

Only the smallest number is kept among equivalence guesses, and the other ones are eliminated. This check is performed as follows.

\begin{enumerate}
\item If a guess contains elements of a set of equivalence numbers $\{a_1, a_2, ..., a_n\} (a_1 < a_2 < ... < a_n) $, check whether they appear in the order as $a_1, a_2, \cdots$ from the most significant digit to the least significant digit.

\item When there exist substitutions that make from a sequence of guesses to the same sequence of guesses, check if it is the smallest among the number of those substitutions applied to the guess. The number of such substitutions is at most $4! = 24$.

\end{enumerate}

By eliminating equivalence guesses, we can fix the first guess to 0123. Similarly, we can reduce the second guesses to 19 guesses: 4567, 0456, 4056, 0145, 0415, 4501, 1045, 1405, 0124, 0142, 0214, 0241, 1204, 1240, 1023, 1032, 0231, 1230. The average number of third and fourth guesses is 1900, which is considerably less than 5040, although the effect is not so great.

\subsection{Computation of sets with 1, 2, or 3 elements}

For sets with 1, 2, or 3 elements, we can find the best guess more quickly. 

\begin{itemize}
\item The best guess for a set with one element is the element itself. In this case, the total number of guesses is one.
\begin{center}
\includegraphics[width=0.9\linewidth]{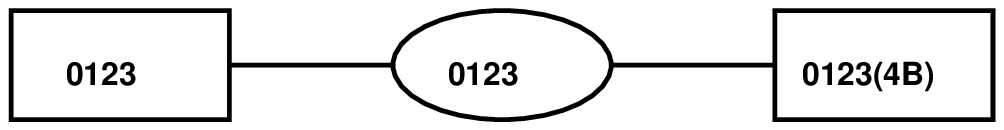} \rule{0cm}{0cm}
\end{center}

\item For a set of 2 elements, the total number of guesses is 3, regardless of which of the two is chosen. This is the best guess.
\begin{center}
\includegraphics[width=0.9\linewidth]{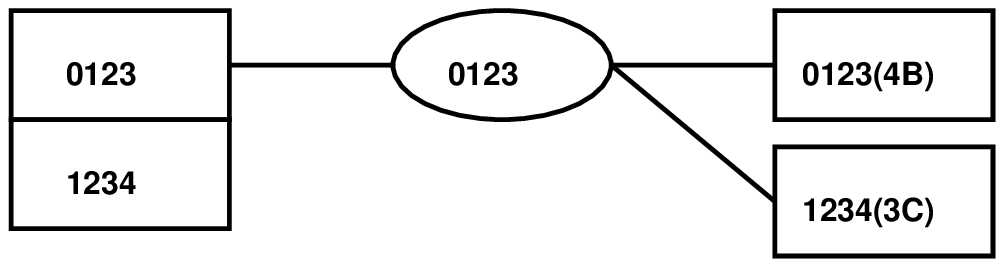} \rule{0cm}{0cm}
\end{center}

\item For a set with three elements, there are two cases. \\If the total number of guesses is 5 when one of the three elements is a guess, then it is the best guess.
\begin{center}
\includegraphics[width=0.9\linewidth]{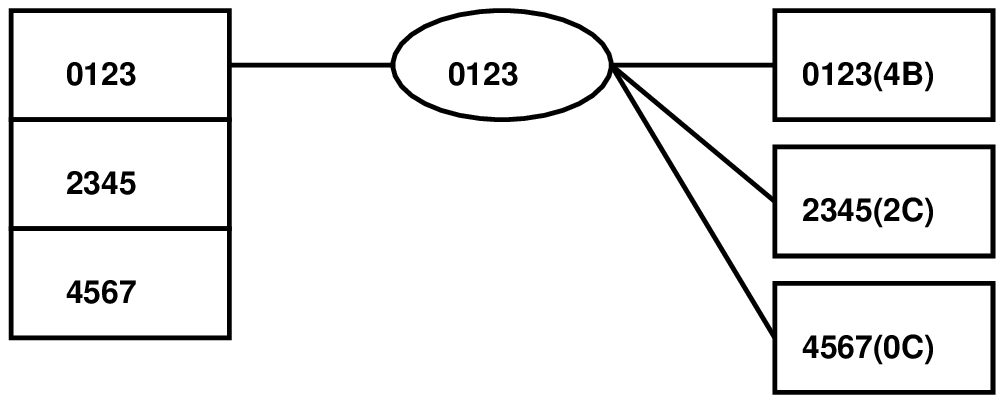} \rule{0cm}{0cm}
\end{center}
When the total number of guesses when one of the three factors is used as a guess is 6 for any of the three factors, the total number of guesses when other guesses are asked is never less than 6, so asking one of the three factors is the best guess to ask.
\begin{center}
\includegraphics[width=0.9\linewidth]{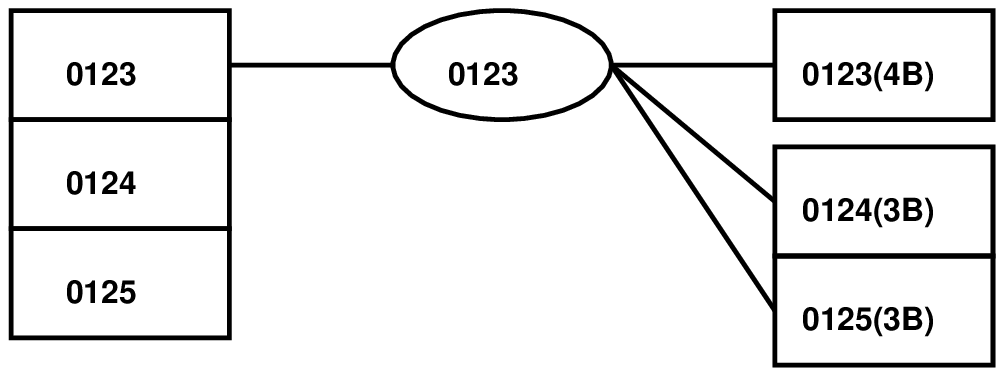} \rule{0cm}{0cm}
\end{center}
\end{itemize}

\subsection{Pruning}

Even when performing an exhaustive search, it is a waste of time to search deeply into clearly suboptimal branches. Therefore, pruning was performed to avoid searching for branches that are not promising.

Suppose that for a given set, the minimum total number of guesses for some guesses has already been computed. When calculating the number of guesses for other guesses, it is meaningless to calculate the total number of guesses unless the total number of guesses is smaller than the minimum total number of guesses at that point in time.

The total number of guesses asked when a guess is asked is equal to or greater than the sum of the expected total number of guesses in the subset obtained by that guess. Therefore, if a value smaller than the sum has already been computed, we know that the guess is not the best guess.

In order to perform this pruning efficiently, the 5040 guesses are sorted by the sum of the expected total number of guesses in the subset at the time the guess is asked, and the branches are searched in order of decreasing value. For this sorting, we use heap sorting. For this purpose, where the number of elements to be extracted is small, heap sort is effective.

%
When the number of elements of a set is determined, we can calculate a lower bound on the total number of guesses from the number of elements. For example, when the number of elements $n$ is $n\leq 14$, the lower limit of the total number of guesses is $1+2(n-1)=2n-1$ because even in the best case, only one element is hit in one guess and $n-1$ elements are hit in two guesses. 

Obtaining the lower bound of the total number of guesses with good accuracy is an important point to make the pruning work effectively. In this case, we have used not only the number of elements in the set but also the types of numbers (4-10) that appear in the set to obtain the lower bound of the total number of guesses. 

In order to find this strategy, we have developed a program that searches for a strategy that maximizes the number of nodes with element one or more up to a depth $m$ for the set of all MOO numbers using $n(4\leq n \leq 10)$ kinds of numbers. For ten types of numbers, the program was able to find the strategy with a depth of 3 in about 80 seconds by taking advantage of symmetry, but for nine types of numbers, the search with a depth of 3 took about 59 hours. The results are shown in Table \ref{max-node}. 

\begin{table}[htbp]
\caption{Maximum number of nodes per depth}
\label{max-node}
\begin{center}
\begin{tabular}{|r|r|r|r|} \hline
kinds of numbers& depth 1 & depth 2 & depth 3 \\ \hline \hline
4 & 4 & 12 & 24 \\ \hline
5 & 8 & 45 & 109 \\ \hline
6 &11 & 78 & 276 \\ \hline
7 &13 &101 &494 \\ \hline
8 &14 &114 &674 \\ \hline
9 &14 &122 &783 \\ \hline
10&14 &127 &864\\ \hline
\end{tabular}
\end{center}
\end{table}

From Table \ref{max-node}, for example, the lower bound of the total number of guesses when there are ten kinds of numbers can be obtained as in Table \ref{min-guess}. 

\begin{table}[htbp]
\caption{The lower limit on the total number of guesses in a set consisting of 10 different numbers}
\label{min-guess}
\begin{center}
\begin{tabular}{|c|c|}  \hline
\# of elements(n)& Total guesses \\ \hline
$ 1 \leq n \leq 14 $   &$ 2n-1 $ \\ \hline
$ 15 \leq n \leq 127 $ & $ 3n-15 $ \\ \hline
$ 128 \leq n \leq 864 $ & $ 4n-142  $ \\ \hline
$ 865 \leq n $ & $ 5n-1006  $ \\ \hline
\end{tabular}
\end{center}
\end{table}

\section{Minimum strategy}

Following the above policy, we developed a program to find the minimum strategy. It is a C language program of about 800 lines. After the first guess is asked, the initial set is divided into 14 groups. You must give a group number (0-13) as an argument so that the program can perform the calculation for the corresponding group. This gives a form of parallelism in which the computation of the 14 sets is performed on different workstations, but this may not have been very useful since the computation time for each set is very different.

We compiled and executed the software on Sparc Station 20 using gcc. As a result, we obtained a minimum strategy with a total computation time of about 27 hours and a total number of guesses of 26274 (average number of guesses is 5.213). The summary of the results is shown in Table {ref{min-strategy}. 

\begin{table}[htbp]
\caption{Summary of the minimum strategy}
\label{min-strategy}
\begin{center}
\begin{tabular}{|c|c|r|r|r|} \hline
First response & Second guess & \# of elements & total guesses & Time(H:M:S) \\ \hline \hline
4B &  & 1 & 0 & 0.0 \\ \hline
2B2C & 0132 & 6 & 15 & 0.0 \\ \hline
1B3C & 0134 & 8 & 22 & 0.1 \\ \hline
4C & 1230 & 9 & 23 & 0.1 \\ \hline
3B & 0245 & 24 & 73 & 0.5 \\ \hline
2B1C & 0145 & 72 & 240 & 2.1 \\ \hline
2B & 0245 & 180 & 659 & 2:08.3 \\ \hline
1B2C & 0245 & 216 & 804 & 1:48.3 \\ \hline
3C & 1435 & 264 & 1004 & 5:00.5 \\ \hline
0C & 4567 & 360 & 1446 & 2.1 \\ \hline
1B & 0456 & 480 & 1913 & 2:52.5 \\ \hline
1B1C & 0245 & 720 & 2992 & 9:28.6 \\ \hline
2C & 1245 & 1260 & 5548 & 2:12:29.7 \\ \hline
1C & 1456 & 1440 & 6495 & 24:24:47.2 \\ \hline
\end{tabular}
\end{center}
\end{table}

In order to give a rough idea of the minimum strategy, which would exceed the number of pages if presented in its original form, the distribution of the number of guesses for 5040 MOOs is shown in Table \ref{bunpu}. 

\begin{table}[htbp]
\caption{Distribution of the minimum strategy}
\label{bunpu}
\begin{center}
\begin{tabular}{|l|r|r|r|r|r|r|r|} \hline
\# of guesses & 1 & 2 & 3 & 4 & 5 & 6 & 7 \\ \hline
counts     & 1 & 7 & 63&  697 & 2424 & 1774 & 74 \\ \hline
\end{tabular}
\end{center}
\end{table}

\section{Strongest strategy}
A small average number of guesses leads to a strong MOO program, but can we say that a strategy with the smallest average number of guesses is the strongest? Of course not. In MOO games, there are only three outcomes: a win, a loss, and a draw, and a win by one move or two moves have the same value. 

In order to examine whether the minimum strategy is the strongest fixed strategy or not, we computed the strategy that has the maximum winning rate when played against the minimum strategy. Let $\gamma_n$ be the expected value of the winning probability against the minimum strategy when guessing a certain number of guesses (a draw is assumed to be 0.5 wins), and let $\Gamma_n = \times 5040 \times 2 - 5040$ be the evaluation value for that number of guesses. This transformation allows us to perform the calculation using only integer arithmetic. This is shown in the table ref{eval-fun}.

\begin{table}[htbp]
\caption{Evaluation Function}
\label{eval-fun}
\begin{center}
\begin{tabular}{|l|r|r|r|r|r|r|r|r|}  \hline
\# of guesses &    1 &    2 &    3 &    4 &    5 &    6 &    7 &    8 \\ \hline \hline
win     & 5039 & 5032 & 4969 & 4272 & 1848 &   74 &    0 &    0 \\  \hline
lose     &    0 &    1 &    8 &   71 &  768 & 3192 & 4966 & 5040 \\ \hline
draw &    1 &    7 &   63 &  697 & 2424 & 1774 &   74 &    0 \\ \hline
Value($\Gamma_n$)   & 5039 & 5041 & 4961 & 4201 & 1080 &-3118 &-4966 &-5040 \\ \hline
\end{tabular}
\end{center}
\end{table}

The program to find the minimum strategy was modified to find the strategy that maximizes the evaluation value (hereafter, the strategy with the highest winning rate). The program took about 58 hours to run. Compared to the case of finding the minimum strategy, it took longer because the evaluation values had to be obtained by multiplication, and the pruning was not as effective as in the case of finding the minimum strategy. A summary of the results is given in Table ref{max-strategy}. 

\begin{table}[htbp]
\caption{Summary of the strategy with the highest winning rate}
\label{max-strategy}
\begin{center}
\begin{tabular}{|c|c|r|r|r|} \hline
First Response & Second guess & Elements & Total guesses & Time(H:M:S) \\ \hline \hline
4B &  & 1 & 0 & 0.0 \\ \hline
2B2C & 0214 & 6 & 16 & 0.5 \\ \hline
1B3C & 0134 & 8 & 22 & 0.5 \\ \hline
4C & 1034 & 9 & 24 & 0.5 \\ \hline
3B & 0456 & 24 & 74 & 1.2 \\ \hline
2B1C & 0145 & 72 & 240 & 17.4 \\ \hline
2B & 0456 & 180 & 661 & 4:01.8 \\ \hline
1B2C & 0245 & 216 & 804 & 3:48.9 \\ \hline
3C & 1435 & 264 & 1004 & 7:29.5 \\ \hline
0C & 4567 & 360 & 1449 & 10.2 \\ \hline
1B & 0456 & 480 & 1915 & 38:18.1 \\ \hline
1B1C & 0145 & 720 & 2996 & 1:46:15.2 \\ \hline
2C & 1245 & 1260 & 5555 & 16:23:42.8 \\ \hline
1C & 1456 & 1440 & 6512 & 38:56:42.9 \\ \hline
\end{tabular}
\end{center}
\end{table}

The average number of guesses for this strategy is 5.221 (26312 total guesses), which is not the minimum guess strategy, but the win rate against the minimum guess strategy is 50.20823 \%. 

There is a highest winning strategy with a winning rate of more than 50 \% against the minimum strategy. Still, it is also possible that there is a strategy with a winning rate of more than 50\% against the highest winning strategy. If there are strategies A, B, and C, there may be no best strategy since the structure is like rock-paper-scissors with $A>B, B>C, C>A$.

However, this possibility was rejected by the experiment. The strategy with the maximum winning rate against the strategy with the highest winning rate was found to be the strategy with the highest winning rate itself, and the winning rate did not exceed 0.5. Therefore, the strategy with the highest winning probability is the strongest strategy.

The distributions of the minimum guess strategy and the strongest strategy are shown in Table \ref{compare}. It can be seen that the strongest strategy has a larger mean value, but more often, the strategy is answered within at most five guesses.

\begin{table}[htbp]
\caption{the minimum strategy and the strongest strategy}
\label{compare}
\begin{center}
\begin{tabular}{|l|r|r|r|r|r|r|r|r|} \hline
\# of guesses & 1 & 2 & 3 & 4 & 5 & 6 & 7 & 8\\ \hline \hline
minimum strategy     & 1 & 7 & 63&  697 & 2424 & 1774 & 74 & 0\\ \hline
strongest strategy  & 1 & 4 & 47&  688 & 2531 & 1628 & 141 & 0\\ \hline
\end{tabular}
\end{center}
\end{table}

\section{verification}

The main body of the program is a program of about 850 lines written in C language, and there is another program of about 650 lines to determine the pruning parameters. Although we have checked the program many times, we cannot deny the possibility that there are bugs in the program. The program and the obtained strategies will be available on the WWW (\url{http://www.ipl.t.u-tokyo.ac.jp/~tanaka/moo/moo.html}
\footnote{The page is at 
\url{https://www.tanaka.ecc.u-tokyo.ac.jp/ktanaka/moo/moo.html}
(in Japanese) and
\url{https://www.tanaka.ecc.u-tokyo.ac.jp/ktanaka/moo/moo-en.html}
(in English)
as of 2022.}
), so if anyone finds any bugs in the program, please let me know. However, we do not plan to offer a prize for bug reports, as Donald E. Knuth does for TeX.

We checked that the obtained results are the correct strategy with a Lisp program written independently of the output of the strategy. However, it is difficult to check whether the obtained strategy is a minimum strategy or not. In the check of the calculation of pi, we compute the result by another program using a different formula and compare the results, but we do not know if it is possible to obtain the result in an acceptable computation time without using techniques such pruning used here. 

\section{Conclusions}

 We have computed the minimum strategy and the strongest strategy for fixed strategies in MOO. However, the strongest strategy is the strongest among the fixed strategies, and dynamic strategies that change moves based on the progress of the opponent's moves may win against the strongest strategy.
For example, the following strategies are considered.
\begin{enumerate}
\item After the first player has guessed the number, the second player chooses a move from the set of solutions.\\

In the minimum strategy and the best strategy, a MOO number that is not included in the set of solutions may be chosen to maximize the evaluation value, but if the first player guesses it, the second player must choose it from the set of solutions otherwise lose the game. 

\item Anticipate the opponent's strategy, predict the number of guesses to be asked until the correct answer at an early stage, and determine your strategy. 
\end{enumerate}
%
However, if the opponent can detect that we have judged $n$ guesses to be correct, he may judge that his strategy includes a solution in the set that can be correct in $n$ times and may guess the number faster. Furthermore, you may change your strategy in anticipation of this. This game may still be difficult.

\end{document}